\documentclass[journal]{IEEE_lsens}
\usepackage{latexsym}
\usepackage{amssymb}
\usepackage{amsmath}
\usepackage[cmintegrals]{newtxmath}
\usepackage{bm}
\usepackage{multicol}
\usepackage{xcolor}
\usepackage{gensymb}
\usepackage{stfloats}
\usepackage{cite}
\usepackage[export]{adjustbox}
\usepackage{caption}
\usepackage{subfigure}
\usepackage{epstopdf}
\usepackage{amsfonts}
\usepackage{amsmath}
\usepackage[dvips]{epsfig}
\usepackage{lipsum,amsmath,multicol} 
\usepackage{svg}
\allowdisplaybreaks[4]	

\begin{document}
\title{Vision Transformer with Convolutional Encoder-Decoder for Hand Gesture Recognition using 24 GHz Doppler Radar}

\author{\IEEEauthorblockN{Kavinda Kehelella\IEEEauthorrefmark{1}, Gayangana Leelarathne\IEEEauthorrefmark{1}\IEEEauthorieeemembermark{1}, Dhanuka Marasinghe\IEEEauthorrefmark{1}\IEEEauthorieeemembermark{1}, Nisal Kariyawasam\IEEEauthorrefmark{1}, Viduneth Ariyarathna\IEEEauthorrefmark{2}\IEEEauthorieeemembermark{2}, Arjuna Madanayake\IEEEauthorrefmark{3}\IEEEauthorieeemembermark{2}, Ranga Rodrigo\IEEEauthorrefmark{1}\IEEEauthorieeemembermark{3}, and Chamira U. S. Edussooriya\IEEEauthorrefmark{1,3}\IEEEauthorieeemembermark{2}}%
\IEEEauthorblockA{\IEEEauthorrefmark{1}Department of Electronic and Telecommunication Engineering, University of Moratuwa, Moratuwa 10400, Sri Lanka\\
\IEEEauthorrefmark{2}Department of Electrical and Computer Engineering, Northeastern University, Boston, MA 02120, USA\\
\IEEEauthorrefmark{3}Department of Electrical and Computer Engineering, Florida International University, Miami, FL 33174, USA \\
\IEEEauthorieeemembermark{1}Student Member, IEEE\\
\IEEEauthorieeemembermark{2}Member, IEEE\\
\IEEEauthorieeemembermark{3}Senior Member, IEEE}%

}


\IEEEtitleabstractindextext{\relax
\begin{abstract}
Transformers combined with convolutional encoders have been recently used for hand gesture recognition (HGR) using micro-Doppler signatures. We propose a vision-transformer-based architecture for HGR with multi-antenna continuous-wave Doppler radar receivers. The proposed architecture consists of three modules: a convolutional encoder-decoder, an attention module with three transformer layers{,} and a multi-layer perceptron. The novel convolutional decoder helps to feed patches with larger sizes to the attention module for improved feature extraction. Experimental results obtained with a dataset corresponding to a two-antenna continuous-wave Doppler radar receiver operating at $24$ GHz (published by Skaria \textit{et al.}) confirm that the proposed architecture achieves an accuracy of $98.3\% $ which substantially surpasses the state-of-the-art on the used dataset.     
\end{abstract}
\vspace{-1ex}
\begin{IEEEkeywords}
Vision transformers, attention mechanisms, deep learning, micro-Doppler signatures, hand gesture recognition.
\vspace{-1ex}
\end{IEEEkeywords}}
\maketitle

\section{Introduction}
\label{sec:int}
\IEEEPARstart{H}{and} gesture recognition (HGR) plays a vital role in human computer interactions, augmented/mixed reality, and human-machine teaming, where specific gestures made by the human hands may be used to control electronics systems. Examples include interfaces to smart phones, vending machines, drones/robots, and gaming devices~\cite{Ahm2021a,Nir,Fu}. On-body device-based approaches, where a person wears or carries a device (e.g., inertial sensors, or radio frequency (RF) identification tags) have been employed for HGR. Device-free approaches such as computer vision, acoustic sensing, and RF sensing have also been employed~\cite{Nir,Fu}. Compared to device-based sensing, device-free sensing is user friendly and widely adopted~\cite{Nir,Fu}. Vision-based and acoustic-based device-free approaches are vulnerable to environmental conditions such as light intensity, rain, smoke, and external noise, while suffering from privacy issues. On the contrary, RF sensing using radar is not as vulnerable to environmental conditions and do not significantly violate privacy. Compared to WiFi-based sensing that {operates} at $2.4$ GHz or $5.8$ GHz, radar-based sensing that operates in millimeter waves (e.g., at $60$ GHz) can detect very small movements of a hand/finger~\cite{Fu,lien2016soli}.

Sensing with continuous-wave (CW)~\cite{skaria,zhao} and frequency-modulated CW radars~\cite{molchanov,Kern2020,Ni,Nin2021} predominantly {utilizes} micro-Doppler signatures for HGR or human activity recognition. Dynamics of a moving object induce Doppler modulation on the {reflected} signal when an RF signal strikes the object in motion~\cite{chen}. CW radars capture micro-Doppler signatures without range information whereas frequency-modulated CW radars capture both micro-Doppler signatures and range information. Recent works~\cite{skaria,Nin2021,Pra2021} demonstrated that radars with multi-antenna receivers achieve higher accuracy than single-antenna radar receivers for HGR applications. 

Radio-frequency machine learning has been utilized to recognize hand gestures with micro-Doppler signatures, where pseudo images generated from received RF signals (e.g., spectrograms and time-Doppler maps) were used as the input~\cite{Ahm2021a}. In~\cite{molchanov,skaria}, convolutional neural networks (CNNs) and in~\cite{Zha2021}, auto-encoders with long short-term memory (LSTM) were employed for HGR. With the popularity of attention-based models with transformers, first employed in natural language processing~\cite{vaswani} and subsequently adopted to computer vision tasks~\cite{Alexander}, recent works on HGR exploited CNNs together with transformers. In~\cite{jaswal2021}, a deep residual three-dimensional CNN with a transformer network was used for HGR with a dataset from \cite{lien2016soli}. A CNN with one-dimensional convolution/correlation and attention-based network was used in~\cite{Lai2021} to classify human activities. In~\cite{haz2019}, an attention+CNN approach via an LSTM was used to recognize human gestures performed from a distance. A CNN feature extractor with an attention-based network was used in \cite{he2019} for person and activity recognition. In~\cite{Zhe2021}, a CNN and a vision-transformer were used for HGR for in-vehicle environments. In these transformer networks, the output feature map of the CNN was directly fed as the input to the transformer. Due to the lower spatial size of the output feature map compared to the input pseudo image to the CNN, the direct feeding leads to patches with lower spatial sizes in the transformers; however, such patches with lower spatial size may hinder the full-potential of transformers~\cite{Alexander}.  

In this paper, we propose a vision-transformer-based architecture for HGR using multi-antenna CW radar. The architecture consists of three modules: 1) a convolutional encoder-decoder, 2) an attention module with three transformer layers, and 3) a multi-layer perceptron (MLP). Compared to previous works on transformers, our architecture employs \emph{a convolutional decoder} to up-sample the output feature map of the convolutional encoder before feeding to the vision transformer. This enables us to use a relatively large patch sizes in the vision transformer {as well as to train with relatively small datasets.} We employ the dataset from~\cite{skaria}, where a two-antenna CW Doppler radar receiver was employed, for validating our algorithms with experiments. The proposed architecture achieves an accuracy of $98.3\% $ which {substantially surpasses the accuracy achieved in~\cite{skaria}.}

\section{Proposed Transformer Architecture}
\begin{figure*}[!t]
    \centering
    \includegraphics[width=1.8\columnwidth]{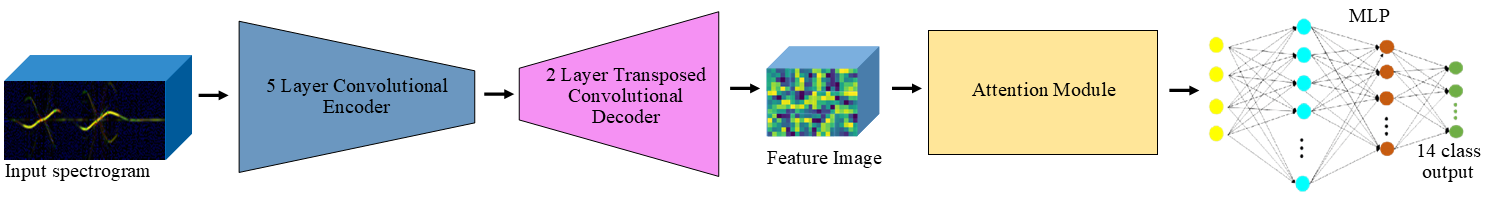}
    \caption{Proposed architecture with a five layer convolutional encoder, a two layer convolutional decoder, an attention module with three vision-transformer layers,  and an MLP for classification.} \label{fig:Proposed architecture}
    \vspace{-1ex}
\end{figure*}

The proposed architecture, shown in Fig.~\ref{fig:Proposed architecture}, consists of a convolutional encoder-decoder, an attention module, and an MLP. We consider a CW radar with one transmit antenna and two receive antennas together with a coherent mixer with in-phase and quadrature. The dataset contains $14$ gestures. See~\cite{skaria} for a system overview and more details on the dataset. The input to the model is a three-channel RGB image generated from spectrograms from two receiver antennas and the angle of arrival of signals obtained from the phase difference between two receiver antennas~\cite{skaria}.

The convolutional encoder consists of five CNN layers, each followed by a max pooling layer. The CNN layers learn features required for subsequent processing with the attention module. Furthermore, the max pooling layers reduce noise in the input and downsample feature maps~\cite{skaria}. The decoder is used to increase the spatial size of the output feature map from the encoder because the size of the output feature map is too small to generate patches for the attention module. The extracted features from the convolutional encoder-decoder are divided into patches, added with positional embeddings~\cite{vaswani,Alexander} and given as the input to the attention module. The attention mechanism allows the modeling of dependencies regardless of their distance in the input or output sequences \cite{vaswani}. We employ three transformer layers in the attention module. These transformer layers are developed using the vision-transformer models in \cite{Alexander}, where multi-head attention is used in each layer. The MLP is employed as the classifier. Next, we describe each module in detail.

\subsection{Convolutional Encoder-Decoder Architecture}
Convolutional encoder-decoder is primarily used to extract features from input images. The number of filters (kernels) and the size of kernels of the five convolutional/correlation layers are presented in Table~\ref{Encoder}. The size of the input is $180\times 60\times 3$, and the input is normalized to the range $[0,1]$ before feeding to the encoder for faster convergence of the model. Rather than feeding the feature maps obtained from the encoder having a size $6\times 2\times 64$ directly to the attention module, we use a convolutional decoder with transposed convolution to up-sample the feature vectors. The convolutional decoder is used to enhance the spatial size of feature maps. There are two transposed convolutional layers with $64$ filters with $1 \times 1$ {kernels} and stride $2 \times 2$. The resultant feature map from the convolutional decoder of size $24\times 8\times 64$ is fed to the attention module.
. 

\begin{table}[!t]
\begin{center}
\captionof{table}{Specifications of the convolutional layers in the encoder.}
\label{Encoder}
\begin{tabular}{ lccc }
\hline
Layer & {Kernel} size & Number of filters & Output size \\ \hline
Input & - & - & $180 \times 60 \times 3$ \\ \hline
Conv $1$ & $7 \times 7$ & $4$ & $180 \times 60 \times 4$ \\ 
Conv $2$ & $5 \times 5$ & $8$ & $90 \times 30  \times 8$ \\ 
Conv $3$ & $3 \times 3$ & $16$ & $45 \times 15 \times 16$ \\ 
Conv $4$ & $3 \times 3$ & $32$ & $23 \times 8  \times 32$ \\ 
Conv $5$ & $3 \times 3$ & $64$ & $12 \times 4  \times 64$ \\
\hline
\end{tabular}
\end{center}
\vspace{-2ex}
\end{table}

\subsection{Attention Module}
\begin{figure*}[t]
    \centering
    \includegraphics[width=1.6\columnwidth]{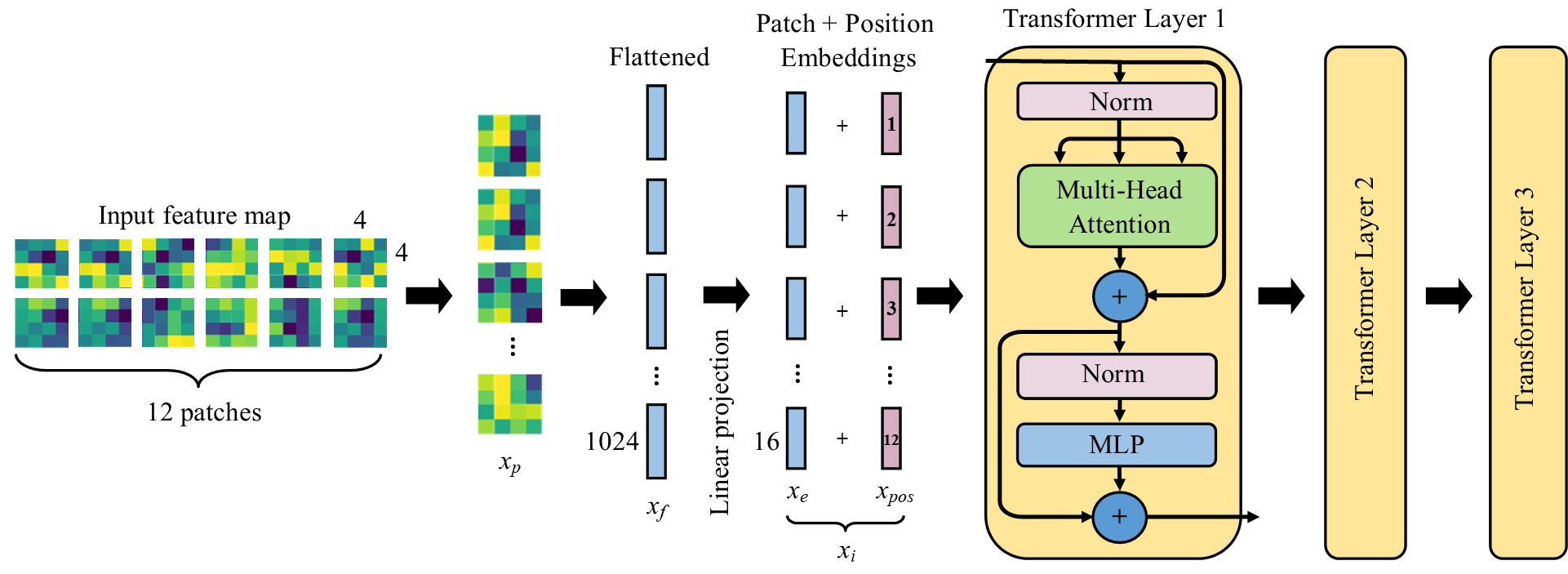}
    \vspace{-1ex}
    \caption{Attention module with patch and positional embeddings and three transformer layers.} \label{fig:Attention Module}
    \vspace{-2ex}
\end{figure*}

In the attention module, shown in {Fig.}~\ref{fig:Attention Module}, the input feature map $\mathbf{x_u} \in \mathbb{R}^{8 \times 24 \times 64}$ is divided into a sequence of square patches $\mathbf{x_p} \in \mathbb{R}^{N \times (p^2\times64)}$, where $N$ is the number of patches and $p$ is the height and width of a patch. Note that $N$ can be calculated as $N = 8\times24/ p^2$. We use $p=4$, and as a result, the $\mathbf{x_u}$ is divided into $12$ patches. Then the tensor $\mathbf{x_p}$ is flattened to produce a tensor $\mathbf{x_f} \in \mathbb{R}^{12 \times 1024}$, which is subsequently projected linearly by a fully-connected layer to produce $\mathbf{x_e} \in \mathbb{R}^{12 \times 16}$ patch embeddings. {Next,} patch embeddings are added with positional embeddings $\mathbf{x_{pos}} \in \mathbb{R}^{12 \times 16}$ to integrate the positional information~\cite{vaswani,Alexander} to generate input embeddings $\mathbf{x_i} \in \mathbb{R}^{12 \times 16}$. The sequence of input embeddings then serves as the input to the vision transformer, where $\mathbf{x_i}$ is normalized and fed into the multi-head attention module. Here query $\mathbf{Q} \in \mathbb{R}^{12 \times 16}$ , key $\mathbf{K} \in \mathbb{R}^{12 \times 16}$ and value $\mathbf{V} \in \mathbb{R}^{12 \times 16}$ matrices are generated by normalizing $\mathbf{x_i}$, i.e., $\mathbf{Q}=\mathbf{K}=\mathbf{V}=Norm(\mathbf{x_i})$.
 
The multi-head attention module contains independent self-attention modules that operate in parallel. Multi-head attention has the advantage of allowing the model to jointly attend {to} information from several representation subspaces\cite{vaswani}. We select the number of heads $h$ as four with the projection dimension $d_k=d_q=d_v$ as 16. The multi-head attention can be performed on $\mathbf{Q}$, $\mathbf{K}$ and $\mathbf{V}$ matrices as
    \begin{align}
    \label{eq:1}
            MultiHead(\mathbf{Q},\mathbf{K},\mathbf{V}) &= Concat\left(\mathbf{H}_1,\mathbf{H}_2,...,\mathbf{H}_j,...,\mathbf{H}_h\right)\mathbf{W}^O, \\
            \mathbf{H}_j &= Attention(\mathbf{Q}\mathbf{W}_j^Q,\mathbf{K}\mathbf{W}_j^K,\mathbf{V}\mathbf{W}_j^V),
    \end{align}
where $\mathbf{W}_j^Q \in \mathbb{R}^{d_k\times \frac{d_k}{h}}$, $\mathbf{W}_j^K \in \mathbb{R}^{d_k,\frac{d_k}{h}}$, $\mathbf{W}_j^V \in \mathbb{R}^{d_k \times \frac{d_k}{h}}$ and $\mathbf{W}^O \in \mathbb{R}^{d_k \times d_k}$ are learnable projection matrices, and $Attention(\hat{\mathbf{Q}},\hat{\mathbf{K}},\hat{\mathbf{V}})$ is defined as 
\begin{align}
    \label{eq:2}
    Attention(\hat{\mathbf{Q}},\hat{\mathbf{K}},\hat{\mathbf{V}}) = softmax\left( \frac{\hat{\mathbf{Q}}\hat{\mathbf{K}}^T}{\sqrt{d_k }} \right)\hat{\mathbf{V}}.
\end{align}
Here, $\hat{\mathbf{Q}}$, $\hat{\mathbf{K}},\hat{\mathbf{V}}$ are projected matrices from $\mathbf{Q}$, $\mathbf{K}$, and $\mathbf{V}$ matrices, respectively~\cite{vaswani,Alexander}. Next, to provide a residual connection to the transformer, the input embeddings and output from the multi-head attention layer are added together. This tensor is then normalized and fed into the MLP block, which consists of two layers with $32$ and $16$ neurons. The output of the transformer layer is fed into the next transformer layer. In our model, we employ three transformer layers $(l)$ as shown in Fig.~\ref{fig:Attention Module}. Finally, the output of the third transformer layer is flattened and fed into the MLP classifier.

\subsection{Multi-Layer Perceptron}
We use an MLP as the classifier with categorical cross-entropy loss function, which is typically employed for multi-class classification problems. The MLP consists of two fully-connected layers followed by a softmax layer with $14$ units for the $14$ gestures. Fully-connected layers have $1024$ neurons and $512$ neurons with $0.5$ dropout to reduce overfitting. Softmax layer calculates corresponding class probabilities for every class. Then the class labels ($y$) can be predicted by performing $\arg \max(\cdot)$ function on the softmax output that returns the index corresponding to the largest probability from the output class probabilities.

\label{sec:trans} 

\section{Experimental Results}
\label{sec:res}

\subsection{Dataset and Training}
\label{ssec:data}
We use the dataset in \cite{skaria} for experimental validation. This dataset was collected using Infineon radar development board BGT$24$MTR$12$ operating at $24$ GHz with a single transmit antenna and an array of two receiving antennas with each receiver producing in-phase and quadrature components. The dataset consists of $14$ hand gestures: (1) single blink, (2) double blink, (3) single push-pull, (4) double push-pull, (5) single round, (6) double round, (7) single swipe, (8) double swipe, (9) single thumbs up, (10) double thumbs up, (11) single waving, (12) double waving, (13) single slide, and (14) double slide, each having $250$ samples. Each sample is a three-channel image of size $180\times60$, where the first two channels are the spectrograms generated from the signals received by two antennas, and the third channel contains the angle of arrival matrix which was generated from the phase difference between the received signals.

We use  $80\%$ of the samples for training and $20\%$ of the samples for testing. {The validation accuracy is obtained from the five-fold cross validation on the training dataset. Moreover, the test accuracy is obtained by independently training the model five times. We present the averages and the standard deviations of test accuracies in the next subsection.} The weighted adaptive moment (AdamW) optimizer is used for training with a learning rate of $0.001$ and a weight decay of $0.0001$. We train the model for $100$ epochs with a batch size of $64$. The transformers allow for substantially higher parallelization, and hence we used an NVIDIA Tesla T$4$ GPU to train the model. 

\vspace{-1.5ex}
\subsection{Experimental Results}
\label{ssec:res}

\subsubsection{Optimum Parameters of the Attention Module}

{The parameters and hyper-parameters of the proposed architecture are tuned to achieve the best validation accuracy.} The height and width of a patch $p$ and the number of dimensions $d_k$ in the linear projection play a critical role in complexity and performance of the proposed architecture. We analyze the performance of the attention module with respect to these two parameters under three criteria as presented in Table~\ref{tab:t3}. Here, we change $p$ and the number of dimensions in the linear projection with fixed parameters for the convolutional encoder-decoder and three transformer layers. We observe that our model with $p=4$ and $d_k=16$ achieves the best accuracy of $98.3\%$. Furthermore, our model has the least number of parameters, with a $25\%$ reduction compared to the next best model. It is evident that attention model with $p=4$ and and $d_k=16$ is the \emph{best among the potential candidates}.

\begin{table}[t!]
\begin{center}
\captionof{table}{Variation of the test accuracy {(avg \textpm \text{} std)}, {the F1-score} and the number of total and trainable parameters with {the patch size ($p$) and the projection dimension ($d_k$)}.}
\vspace{-1.5ex}
\begin{tabular}{ ccrrcc }
\hline
\begin{tabular}[c]{@{}l@{}}{$p$}\end{tabular} & \begin{tabular}[c]{@{}l@{}} {$d_k$}\end{tabular} & \begin{tabular}[c]{@{}l@{}}Total \\ parameters\end{tabular} & \begin{tabular}[c]{@{}l@{}}Trainable \\ parameters\end{tabular} & \begin{tabular}[c]{@{}l@{}}Test \\ accuracy ($\%$)\end{tabular} &
\begin{tabular}[c]{@{}l@{}}{F1-score (\%)} \end{tabular} \\ \hline
2   & 16    & 1,375,190     & 1,374,942     & 95.2 \textpm {0.55} & {95.1} \\ 
2   & 32    & 2,213,574     & 2,213,326     & 95.8 \textpm {0.65} & {95.8} \\ 
2   & 64    & 3,982,502     & 3,982,254     & 94.9 \textpm {0.70} & {94.7}\\ 
\textbf{4}   & \textbf{16}    & \textbf{797,078}       & \textbf{796,830}       & \textbf{98.3} \textpm {\textbf{0.50}} & {\textbf{98.4}} \\ 
4   & 32    & 1,057,350     & 1,057,102     & 97.7 \textpm {0.58} & {97.5}\\ 
4   & 64    & 1,670,054     & 1,669,806     & 95.6 \textpm {0.39} & {95.5}\\
\hline
\end{tabular}
\label{tab:t3}
\end{center}
\vspace{-3ex}
\end{table}

The number of transformer layers $l$ in the attention module is the other important parameter. A model with more transformer layers has the ability to extract more informative features, however, at the same time, {the} model complexity increases. Moreover, more transformer layers tend to increase the variance of the model because of the relatively small dataset that was used. We analyzed the performance of our model by varying $l$ from $1$ to $6$ while fixing other parameters, and the results are presented in Table~\ref{transformer_layers}. We can see that \emph{the best accuracy of $98.3\%$ is achieved with $l=3$}, which is selected for our architecture, even though the number of total and trainable parameters are slightly higher compared to $l=1$ and $l=2$.

\begin{table}[t!]
\begin{center}
\captionof{table}{Variation of the test accuracy {(avg \textpm \text{} std)}, {the F1-score} and the number of total and trainable parameters with {the number of transformer layers ($l$).}}
\vspace{-1.5ex}
\begin{tabular}{ccccc}
\hline
  \begin{tabular}[c]{@{}l@{}}{$l$} \end{tabular} &
  \begin{tabular}[c]{@{}l@{}}Total \\ Parameters\end{tabular} &
  \begin{tabular}[c]{@{}l@{}}Trainable \\ Parameters\end{tabular} &
  \begin{tabular}[c]{@{}l@{}}Test \\ accuracy (\%)\end{tabular} &
  \begin{tabular}[c]{@{}l@{}}{F1-score (\%)}\end{tabular} \\ \hline
1 & 786,198 & 785,950 & 96.0 \textpm {0.75} & {96.1}         \\ 
2 & 791,638 & 791,390 & 94.5 \textpm {0.79} & {94.3}        \\ 
\textbf{3} & \textbf{797,078} & \textbf{796,830} & \textbf{98.3} \textpm {\textbf{0.50}} & {\textbf{98.4}}\\ 
4 & 802,518 & 802,270 & 97.5 \textpm {0.53} & {97.4}        \\ 
5 & 807,958 & 807,710 & 97.3 \textpm {0.75} & {97.5}         \\ 
6 & 813,398 & 813,150 & 96.9 \textpm {0.67} & {96.9}         \\ \hline
\end{tabular}
\label{transformer_layers}
\end{center}
\vspace{-3ex}
\end{table}

\subsubsection{Ablation Study}
Our architecture is composed with both convolutional encoder-decoder  and an attention module. Since the state-of-art deep neural network model for this dataset is a CNN, we select the CNN in our model as the baseline. We employ an ablation study to verify the improved performance with the convolutional decoder and the attention module. We also trained the standalone modules (decoder and attention module) on the dataset and the results are presented in Table \ref{CNN and transformer}.

When training with standalone attention module, three-channel images are split into patches with $p=36$ and provided as the input to the encoder. Furthermore, for the combination of the convolutional encoder and the attention module, $p=1$ and $d_k=16$ are selected.

\begin{table}[t!]
\begin{center}
\captionof{table}{Ablation study with different combinations of modules.}
\vspace{-1.5ex}
    \begin{tabular}{ p{4cm}cc}
        \hline
        Architecture & Test accuracy (\%) & {F1-score (\%)}\\
        \hline
        Convolutional Encoder & 96.7 \textpm {0.41} & {96.7}  \\ 
        Convolutional Encoder-Decoder & 96.7 \textpm {0.59} & {96.6} \\
        Attention Module & 91.9 \textpm {0.73} & {91.8} \\
        Convolutional Encoder + Attention Module & 96.0 \textpm {0.67} & {95.9} \\
        \textbf{Convolutional Encoder-Decoder + Attention Module} & \textbf{98.3} \textpm {\textbf{0.50}} & {\textbf{98.4}} \\
        \hline
    \end{tabular}
    \label{CNN and transformer}
\end{center}
\vspace{-2.5ex}
\end{table}

We observed that the standalone convolutional encoder outperformed the standalone attention module by $4.8\ $ {percentage points}. This is because convolution has inductive bias such as translation invariance which lacks in transformers~\cite{Alexander}. Therefore, when trained on small datasets, transformers do not generalize the model. Furthermore, the combination of the convolutional encoder and the attention module (with $p=1$ and $d_k=16$) outperforms the standalone attention module by $4.1\ ${percentage points}. More importantly, our architecture outperforms the four other approaches. In particular, our architecture achieves $2.3\ ${percentage points} higher accuracy than that of the combination of the convolutional encoder and the attention module \emph{verifying the importance of the convolutional decoder}.

\subsubsection{Comparison with Other Architectures}
We compare accuracies achieved with our model, CNN architecture in~\cite{skaria}, and two popular image classification deep neural network models: ResNet50~\cite{resnet} and VGGNet16~\cite{vggnet}, pretrained on ImageNet dataset. We modify the last softmax layer of both ResNet50 and VGGNet16 to have $14$ classes. In transfer learning, we train only the parameters of the last layer of both networks while freezing other layers with pretrained parameters. The achieved test accuracies are presented in {Table} \ref{architectures}. Our model outperforms other three models, in particular, the original work \cite{skaria} by a margin of $2.8$ {percentage points. Similar to other transformer-based architectures, a limitation of the proposed architecture is that the number of total parameters (797,078) is considerably higher than that of the CNN architecture (7676)~\cite{skaria} leading to higher computational and memory complexities. The interference times of the CNN~\cite{skaria}, ResNet50~\cite{resnet}, VGGNet16~\cite{vggnet} and the proposed architectures are 35 ms, 130 ms, 170 ms, and 96 ms, respectively. The inference time of the proposed architecture is higher than that of the CNN architecture~\cite{skaria} and lower than those of the ResNet50~\cite{resnet} and VGGNet16~\cite{vggnet} architectures. Even though the proposed architecture has more layers and parameters compared to the CNN~\cite{skaria}, the inference time is only $\approx\!\!3$ times higher due to the highly parallel implementation of the attention module.} Future work may consider the reduction of the complexity of the proposed architecture using pruning techniques~\cite{Vad2022_b}.

\begin{table}[t!]
\begin{center}
\captionof{table}{{Classification performance achieved with different models.}}
\vspace{-1.5ex}
    \begin{tabular}{ lccccc }
        \hline
        Architecture & Test accuracy & Precision & Recall & F1-score (\%)\\
        \hline
        CNN \cite{skaria}  & {95.1} \textpm {0.54} & {95.0} & {95.1} & {95.1} \\
        ResNet50 \cite{resnet} & 92.1 \textpm {0.62} & {92.0} & {92.1} & {92.0} \\
        VGGNet16 \cite{vggnet}  & 95.0 \textpm {0.61} & {94.8} & {94.7} & {94.7} \\
        \textbf{Our model} &  \textbf{98.3} \textpm {\textbf{0.50}} & {\textbf{98.4}} & {\textbf{98.4}} & {\textbf{98.4}} \\
        \hline
    \end{tabular}
    \label{architectures}
\end{center}
\vspace{-4ex}
\end{table}

\vspace{-1.5ex}
\section{Conclusion}
We propose a vision-transformer-based architecture for HGR with multi-antenna CW Doppler radar receivers. The convolutional decoder up-samples the output feature map of the convolutional encoder enabling us to feed patches with larger sizes to the attention module. This results in improved feature extraction in the attention module. Experimental results confirmed that the proposed architecture achieves an accuracy of $98.3\% $ which is substantially higher than the state-of-the-art on the used dataset, and the ablation study confirms that the convolutional decoder improves the accuracy on HGR.   

\section*{Acknowledgment}
\scriptsize
This work was financially supported in part by the Senate Research Committee, University of Moratuwa under grant SRC/LT/2020/08.

\normalsize
\bibliographystyle{IEEEtran}    
\bibliography{IEEEabrv,Refs}

\begin{thebibliography}{10}
\providecommand{\url}[1]{#1}
\csname url@samestyle\endcsname
\providecommand{\newblock}{\relax}
\providecommand{\bibinfo}[2]{#2}
\providecommand{\BIBentrySTDinterwordspacing}{\spaceskip=0pt\relax}
\providecommand{\BIBentryALTinterwordstretchfactor}{4}
\providecommand{\BIBentryALTinterwordspacing}{\spaceskip=\fontdimen2\font plus
\BIBentryALTinterwordstretchfactor\fontdimen3\font minus
  \fontdimen4\font\relax}
\providecommand{\BIBforeignlanguage}[2]{{%
\expandafter\ifx\csname l@#1\endcsname\relax
\typeout{** WARNING: IEEEtran.bst: No hyphenation pattern has been}%
\typeout{** loaded for the language `#1'. Using the pattern for}%
\typeout{** the default language instead.}%
\else
\language=\csname l@#1\endcsname
\fi
#2}}
\providecommand{\BIBdecl}{\relax}
\BIBdecl

\bibitem{Ahm2021a}
S.~Ahmed, K.~D. Kallu, S.~Ahmed, and S.~H. Cho, ``{Hand gestures recognition
  using radar sensors for human-computer-interaction: A review},'' \emph{Remote
  Sensing}, vol.~13, no.~3, pp. 1--24, Feb. 2021.

\bibitem{Nir}
I.~Nirmal, A.~Khamis, M.~Hassan, W.~Hu, and X.~Zhu, ``{Deep learning for
  radio-based human sensing: Recent advances and future directions},''
  \emph{IEEE Communications Surveys \& Tutorials}, 2021.

\bibitem{Fu}
B.~Fu, N.~Damer, F.~Kirchbuchner, and A.~Kuijper, ``{Sensing technology for
  human activity recognition: A comprehensive survey},'' \emph{IEEE Access},
  vol.~8, pp. 83\,791--83\,820, 2020.

\bibitem{lien2016soli}
J.~Lien, N.~Gillian, M.~E. Karagozler, P.~Amihood, C.~Schwesig, E.~Olson,
  H.~Raja, and I.~Poupyrev, ``{Soli: Ubiquitous gesture sensing with millimeter
  waver radar},'' \emph{ACM Transactions on Graphics}, vol.~35, no.~4, pp.
  1--19, Jul. 2016.

\bibitem{skaria}
S.~Skaria, A.~Al-Hourani, M.~Lech, and R.~J. Evans, ``Hand-gesture recognition
  using two-antenna {D}oppler radar with deep convolutional neural networks,''
  \emph{IEEE Sensors Journal}, vol.~19, no.~8, pp. 3041--3048, 2019.

\bibitem{zhao}
R.~Zhao, X.~Ma, X.~Liu, and F.~Li, ``{Continuous human motion recognition using
  micro-Doppler signatures in the scenario with micro motion interference},''
  \emph{IEEE Sensors Journal}, vol.~21, no.~4, pp. 5022--5034, 2021.

\bibitem{molchanov}
P.~Molchanov, S.~Gupta, K.~Kim, and K.~Pulli, ``{Short-range FMCW monopulse
  radar for hand-gesture sensing},'' in \emph{IEEE Radar Conference}, 2015, pp.
  1491--1496.

\bibitem{Kern2020}
N.~Kern, M.~Steiner, R.~Lorenzin, and C.~Waldschmidt, ``{Robust Doppler-based
  gesture recognition with incoherent automotive radar sensor networks},''
  \emph{IEEE Sensors Letters}, vol.~4, no.~11, pp. 1--4, Nov. 2020.

\bibitem{Ni}
Z.~Ni and B.~Huang, ``{Open-set human identification based on Gait radar
  micro-Doppler signatures},'' \emph{IEEE Sensors Journal}, vol.~21, no.~6, pp.
  8226--8233, 2021.

\bibitem{Nin2021}
A.~Ninos, J.~Hasch, and T.~Zwick, ``{Real-time macro gesture recognition using
  efficient empirical feature extraction with millimeter-wave technology},''
  \emph{IEEE Sensors Journal}, vol.~21, no.~13, pp. 15\,161--15\,170, Jul.
  2021.

\bibitem{chen}
V.~Chen, F.~Li, S.-S. Ho, and H.~Wechsler, ``Micro-{D}oppler effect in radar:
  Phenomenon, model, and simulation study,'' \emph{IEEE Transactions on
  Aerospace and Electronic Systems}, vol.~42, no.~1, pp. 2--21, 2006.

\bibitem{Pra2021}
A.~A. Pramudita, Lukas, and Edwar, ``Contactless hand gesture sensor based on
  array of {CW} radar for human to machine interface,'' \emph{IEEE Sensors
  Journal}, vol.~21, no.~13, pp. 15\,196--15\,208, Jul. 2021.

\bibitem{Zha2021}
B.~Zhang, L.~Zhang, M.~Wu, and Y.~Wang, ``Dynamic gesture recognition based on
  {RF} sensor and {AE-LSTM} neural network,'' in \emph{International Symposium
  on Circuits and Systems}, 2021, pp. 1--5.

\bibitem{vaswani}
A.~Vaswani, N.~Shazeer, N.~Parmar, J.~Uszkoreit, L.~Jones, A.~N. Gomez,
  {\L}.~Kaiser, and I.~Polosukhin, ``Attention is all you need,'' in
  \emph{{Advances in Neural Information Processing Systems}}, 2017, pp. 1--11.

\bibitem{Alexander}
A.~Dosovitskiy, L.~Beyer, A.~Kolesnikov, D.~Weissenborn, X.~Zhai,
  T.~Unterthiner, M.~Dehghani, M.~Minderer, G.~Heigold, S.~Gelly,
  T.~Unterthiner, and X.~Zhai, ``An image is worth 16x16 words: Transformers
  for image recognition at scale,'' in \emph{International Conference on
  Learning Representations}, 2021, pp. 1--21.

\bibitem{jaswal2021}
G.~Jaswal, S.~Srirangarajan, and S.~D. Roy, ``{Range-Doppler} hand gesture
  recognition using deep {residual-3DCNN} with transformer network,'' in
  \emph{International Conference on Pattern Recognition}, 2021, pp. 759--772.

\bibitem{Lai2021}
G.~Lai, X.~Lou, and W.~Ye, ``Radar-based human activity recognition with {1-D}
  dense attention network,'' \emph{IEEE Geoscience and Remote Sensing Letters},
  vol.~19, pp. 1--5, Jan. 2022.

\bibitem{haz2019}
S.~Hazra and A.~Santra, ``Radar gesture recognition system in presence of
  interference using self-attention neural network,'' in \emph{IEEE
  International Conference on Machine Learning and Applications}, 2019, pp.
  1409--1414.

\bibitem{he2019}
Y.~He, X.~Li, and X.~Jing, ``A mutiscale residual attention network for
  multitask learning of human activity using radar {micro-Doppler}
  signatures,'' \emph{Remote Sensing}, vol.~11, no.~21, pp. 1--18, Nov. 2019.

\bibitem{Zhe2021}
L.~Zheng, J.~Bai, X.~Zhu, L.~Huang, C.~Shan, Q.~Wu, and L.~Zhang, ``Dynamic
  hand gesture recognition in in-vehicle environment based on {FMCW} radar and
  transformer,'' \emph{Sensors}, vol.~21, no.~19, pp. 1--20, Sep. 2021.

\bibitem{resnet}
K.~He, X.~Zhang, S.~Ren, and J.~Sun, ``{Deep residual learning for image
  recognition},'' in \emph{IEEE Conference on Computer Vision and Pattern
  Recognition}, 2016, pp. 770--778.

\bibitem{vggnet}
K.~Simonyan and A.~Zisserman, ``{Very deep convolutional networks for
  large-scale image recognition},'' \emph{arXiv preprint arXiv:1409.1556},
  2014.

\bibitem{Vad2022_b}
{Vadera, Sunil and Ameen, Salem}, ``{Methods for Pruning Deep Neural
  Networks},'' \emph{{IEEE Access}}, vol.~{10}, pp. {63\,280--63\,300}, {Jun.}
  {2022}.

\end{thebibliography}

\end{document}